\documentclass{sig-alternate-10pt}

\makeatletter
\let\@copyrightspace\relax
\makeatother

\usepackage{floatrow}
\newfloatcommand{capbtabbox}{table}[][\FBwidth]
\usepackage{blindtext}

\usepackage{graphicx}
\usepackage{epsfig}
\usepackage{epstopdf}
\usepackage{subfigure}
\usepackage{url}
\usepackage{cite}%
\usepackage{color}
\usepackage{balance}
\usepackage{multirow}
\usepackage[normalem]{ulem}
\usepackage{fmtcount}

\usepackage{times}
\usepackage{cite}
\usepackage{amsfonts,amssymb,amsmath}
\usepackage{balance}
\usepackage{algorithm}
\usepackage{algorithmic,eqparbox,array}

\renewcommand{\paragraph}[1]{\textbf{#1: }}
\newif\ifdebugdoc\debugdoctrue

\ifdebugdoc
\newcommand{\fyi}[1]{\footnote{\textcolor{blue}{fyi:#1}}}

\newcommand{\del}[1]{\textcolor{blue}{\sout{#1}}}
\newcommand{\outline}[1]{\textbf{\colorbox{yellow}{Outline:}\textcolor{red}{#1.}}}

\newcommand{\chunyi}[1]{\footnote{\colorbox{yellow}{Chunyi:} #1.}}

\else
\newcommand{\fyi}[1]{}

\newcommand{\del}[1]{}
\newcommand{\chunyi}[1]{}
\newcommand{\outline}[1]{}
\fi

\begin{document}
%\title{Open the Black Box: The devil is in your cell phone bill}
%\title{Open the Black Box: Look into Mobile Data Accounting in Cellular Networks}
%\title{The devil is in your cell phone bill}
\title{A Survey Report on Hardware Trojan Detection by Multiple-Parameter Side-Channel Analysis}

\author{Samir R Katte$^1$,Keith E Fernandez$^1$\\
UCLA Electrical Engineering, Los Angeles, CA 90095$^1$\\
}

\maketitle

\begin{sloppypar}

% edit version

\section{Abstract}

\parskip 0.1in
A major security threat to an integrated circuit (IC) design is the Hardware Trojan attack which is a malicious modification of the design. Previously several papers have investigated into side-channel analysis to detect the presence of Hardware Trojans. The side channel analysis were prescribed in these papers as an alternative to the conventional logic testing for detecting malicious modification in the design. It has been found that these conventional logic testing are ineffective when it comes to detecting small Trojans due to decrease in the sensitivity due to process variations encountered in the manufacturing techniques. The main paper \cite{1} under consideration in this survey report focuses on proposing a new technique to detect Trojans by using multiple-parameter side-channel analysis. The novel idea will be explained thoroughly in this survey report. We also look into several other papers \cite{2}\cite{3}\cite{4}\cite{5}\cite{6}, which talk about single parameter analysis and how they are implemented. We analyzed the short comings of those single parameter analysis techniques and we then show how this multi-parameter analysis technique is better. Finally we will talk about the combined side-channel analysis and logic testing approach in which there is higher detection coverage for hardware Trojan circuits of different types and sizes. 

\vspace{0.1cm}
\begin{flushleft}
\textbf{\textit{Keywords}}: Hardware Trojans, multiple parameter side-channel analysis, process variation, logic testing.
\end{flushleft}

\section{Introduction}
\parskip 0.1in
Hardware Trojans are the most recent security threats to integrated circuits (ICs) and it is necessary to check whether a given IC does not have any malicious modification i.e. a hardware Trojan is not inserted into it. One major cause of such attacks is due to outsourcing of the IC fabrication to foreign countries. The foundry if untrusted, it can insert a hardware Trojan in the existing design of the IC. The adversary should be intelligent enough to insert a Trojan which is undetected by conventional testing techniques but it gets activated when the IC is being used for its original operation. There are two ways to create such a Trojan. One way is to trigger its operation externally while the other way is to make the Trojan dependent on rare circuit conditions.

The author of the main paper under consideration \cite{1} uses some terminologies which are important to understand the analysis of the paper. The node affected by the Trojan is called the payload. The condition at which the Trojan is activated has been called as the trigger condition. This trigger condition can be purely combinational or sequentially related to the clock or a set of rare events. The malicious effects of Trojan payloads can be either a passive effect like revealing some secret information in a cryptographic IC or an active effect like entirely changing the functionality of the circuitry.

There are two types of non-destructive Trojan detection techniques: logic testing and side-channel approaches.  The conventional testing of an IC aims at the functional validation and it does not provide the high coverage for Trojan detection. So statistical logic testing has been suggested before which generates the structural tests to activate rare events and propagate the malicious effects to the primary outputs. This approach can be used when the size of the Trojan is ultra small i.e only of few gates in size. It becomes cumbersome if the kind of Trojan is complex sequential or if the number of Trojan instances is large.

Apart from the Trojan detection technique mentioned above, there are several other physical side-channel parameters like the power signature which can be measured to detect the presence of hardware Trojans. Unlike in the statistical logic testing method, these approaches do not require us to trigger the Trojan and observe its impact at the primary output. But still there can be extreme variations in the measured side-channel parameters due to process variations. 

The present side channel approaches have certain shortcomings. Firstly with the increase in the process variations, the process calibration techniques and hence the Trojan detection sensitivity reduces. Secondly they consider only die-to-die process variation and ignore local within-die variations. Lastly they need design modifications which can be exploited by the adversary. Also as the size of the circuit increases and the size of the Trojan increases, it becomes tougher to detect Trojans because the detection sensitivity of the side-channel approaches decreases.

The paper\cite{1} describes a novel non-invasive multiple parameter side-channel analysis approach for effective detection of complex Trojans under large process-induced parameter variations. The method looks at the correlation of the intrinsic leakage (I$_{DDQ}$) to the maximum operating frequency (F$_{MAX}$), so that we can identify fast, intrinsically leaky ICs from the defective ones. This method does not only look at the power signature. It uses the dependencies between the transient supply current (I$_{DDT}$) and F$_{MAX}$ of the circuit to find the Trojan infected ICs. 

There are several salient features of this paper as it touches upon this novel method. The technique described requires zero modification to the design flow and there is no hardware overhead. The major contribution of the paper is that it provides a theoretical analysis of the relationship between the multiple parameters and how it is used for reducing the process noise and for identifying Trojans. The FPGA based approach described provides both simulation verification and hardware validation. The impact of this paper is immense because it is the one of the first few papers to look into multiple parameter side-channel analysis, unlike previous papers which looked into single parameters only. Firstly it looks into a structural test-generation approach which minimizes the switching activity in different parts of the design and increases the activity of an arbitrary Trojan in the region under test. After that it proposes using power gating techniques to improve signal to noise by reducing the background current. It also introduces a third parameter called quiescent current or I$_{DDQ}$ to improve the confidence of detection. It also looks into how to increase the detection sensitivity by proper choice of test conditions like operating voltage and frequency. Subsequently the paper also proposes the integration of the proposed side-channel approach and the statistical logic testing approach, which can detect Trojans of different types and sizes.

The rest of this survey paper is organized as follows: Section 3 talks about the related work of this topic which basically talks about the work done in the papers apart from the main paper under survey. Section 4 explains the theory behind the operation of hardware Trojans and other technical terms. Section 5 compares the main paper with other papers and presents our findings. Section 6 gives an overview of the results obtained in the main paper. Finally Section 7 concludes the paper and section 8 is our critique of the entire set of papers. Also we list the references and the list of papers presented in the class by the authors in section 9 and section 10 respectively.

\section{Related Work}
\parskip 0.1in
There have been several previous attempts to detect hardware Trojans in fabricated ICs. Most of these papers talk about a single parameter side-channel analysis to detect the presence of a malicious modification in the design. 

\cite{2} investigates a power supply transient signal analysis method for detecting Trojans. It basically analyzes multiple power port signals. More precisely the paper\cite{2} focuses on determining the smallest detectable Trojan in a set of process  simulation models. Ten different layouts are being analyzed here. One of them is Trojan free and the others are inserted with a few gates to model a Trojan. Simulated models are extracted from the layers and the simulation data is analyzed to know when a Trojan can be detected.  The results of the sensitivity analysis show that it is possible to reliably detect unactivated Trojans which are created using as few as four standard cell gates.

\cite{3} focuses on hardware Trojan detection using path delay fingerprint. The authors specially focus on the detection of explicit payload Trojan. This paper introduced this new category of hardware Trojans based on how the payload part of the Trojans works. The new category is divided into implicit Trojans and explicit Trojans. The explicit payload Trojan works under a typical two-phase manner: triggered and propagated payload. When the Trojan is triggered, the payload part will change the internal control signals or data signals and we result in the chip to perform erroneous or propagate secret information such as symmetrical keys to some output pins. This type of Trojan will insert extra delay in some paths passing those signals. On the other hand, implicit payload Trojans has a similar trigger part as the explicit payload Trojan but different payload working mechanics. The implicit payload Trojan does not compromise internal signals but only takes these signals as a stimulus of the trigger. After the Trojan is triggered, the implicit payload part will behave in a different way than it does in explicit Trojan. The implicit Trojan can leak secret information by emitting radio waves or may destroy the entire chip. The signal to trigger the implicit payload Trojan has a larger capacity load and hence consumes more power and it also makes some path delays larger. But if we compared the path delay of the explicit payload Trojan, the added delay by the implicit payload Trojan is smaller and much harder to detect. Another way to distinguish between these two types of Trojans is that an explicit payload Trojan can be detected using exhaustive traditional functional tests unlike implicit payload Trojan. \cite{3} had experimental results which showed the detection rate of explicit payload Trojans to be 100\%. The method described by the authors should be developed further to be used for implicit payload Trojan detection.  

\cite{4} talks about single parameter side-channel analysis wherein it uses the transient power analysis of the IC for Trojan detection. The paper proposes a non-destructive approach which characterizes and compares transient power signature using principle component analysis. The approach is validated with hardware measurement results. The test setup is FPGA based and this approach can discover small (<1.1\% area) Trojans under large noise and variation.

\cite{5} discusses a technique for precisely measuring the combinational delay of an arbitrarily large number of register-to-register paths internal to the functional portion of the IC. It is suggested that this technique can be used to provide the desired authentication and design alteration detection. This technique is low cost and it does not affect the main IC functionality and can be performed at-speed at both the test-time and run time. Understanding the working of a Physical Unclonable Functions (PUF) is important for completely understanding the contents of \cite{5}. The theory of PUF is covered at the end of the next section. In \cite{5}, the authors suggested the technique which can be used to generate longer signatures than the existing PUF designs. Also unlike other techniques that are applied to non-functional paths specifically inserted to be used as a PUF, this technique can be performed on the functional paths of the core circuit without affecting timing and functionality. Thus it makes this technique significantly more difficult for an attacker to bypass, remove, or spoof the signature extraction unit.

A novel on-chip structure including a ring oscillator network (RON), distributed across the entire chip, is proposed in \cite{6} to verify whether the chip is Trojan-free or not. This structure eliminates the problem of measurement noise and localizes the measurement of dynamic power. The authors have presented the simulation results for Trojans inserted into 90nm technology circuits. There are also some experimental results present in the paper which demonstrates the efficiency and the scalability of the RON architecture for Trojan detection.
The next section talks about the theory required to understand the analysis of the papers.

\section{Theory}
\parskip 0.1in
Hardware Trojans are classified based on the method of activation and the effect it has on the circuit functionality. If the Trojan is combinationally triggered, for example an A=B condition occurs at the trigger inputs of the Trojan which changes the expected output at the node from ER to an incorrect value ER*. The adversary chooses a very rare condition for the Trojan activation so that it is not triggered during conventional manufacturing tests. Sequentially triggered Trojans are activated by occurrence of a sequence of rare events or after a period of continuous operation. An example of sequentially triggered Trojans is that an asynchronous k-bit counter activates a Trojan when the count reaches 2$^{k}$-1 and hence modifies the node ER to an incorrect value of ER*. Another type of Trojan is one which consists of a linear feedback shift register. It is used to leak the secret key in cryptographic hardware by helping in side-channel attacks.

It is difficult to have a single Trojan detection mechanism because of the variety of Trojans that need to be detected. Destructive testing of a chip is expensive due to depackaging, de-metallization and micro-photography implemented in the whole process. Also this technique is not feasible because the attacker can insert the Trojan into a small subset of the manufactured IC. So non-destructive methods are used. The logic based testing aims at triggering rare events at the internal nodes of the circuit to activate Trojans and then observe the outputs. Meanwhile in side-channel analysis-based Trojan detection, we observe the effect of Trojan insertion on the parameters like circuit transient current, leakage current, etc. If the values are beyond a particular threshold value, we can say that a Trojan is detected. 

But both the Trojan detection techniques have their own pros and cons. The main problem in logic testing is the extremely large Trojan design space and hence the complete enumeration and the test generation are infeasible in computational terms. Side-channel analysis is advantageous because it does not look into the malfunction of the circuit but looks into the changes in the side-channel parameter. Problems in side-channel analysis include large process-induced parameter variations and measurement noise which can mask the effect of insertion of small Trojans. But none of the previous works suggest techniques which eliminates the local and the large process variations. 

\cite{5} talks about Physical Unclonable Functions (PUF). So we will explain PUF briefly so that it will be easier to understand the technique proposed by the author. PUFs are functions that map a set of challenges to responses that are generated from, and hence reflect, the unique physical characteristics of each device. Therefore PUFs can provide higher security than other soft-key-based cryptographies because they extract the security information from the physical system itself, rather than storing this information in non-volatile memory. The working of a PUF is explained as followed: An n-bit challenge vector (say C) is input to an n-stage string of switch blocks as the control vector i.e. one bit for each switch block. Each switch block has two inputs, two outputs and a control bit. Depending on the value of the control bit, the inputs of the switch block will either go directly to the outputs or will be exchanged. Thus the circuit can create a pair of delay paths for each input vector. For evaluation, a rising edge is input to stimulate both of the circuit paths. The signal passes through these two paths and one of the outputs of the last switch block is used as the response to that challenge vector. The known output to the challenge is compared to the response for authentication.

\section{Comparison with others papers}
\parskip 0.1in
The main issue with single side channel techniques is that they are prone to large process variations and it has a large susceptibility to process noise. Most Side Channel Techniques described in papers \cite{2}, \cite{3},  \cite{4},  \cite{5}  and  \cite{6} could only deal with one of the two problems. But as scaling increases, these problems are more prominent at lower design nodes. A case may arise where a single parameter signature of a Trojan-Infested IC falls within the threshold of a Genuine IC. 

This paper \cite{1} deals with a unique multiple side parameter method which models process variation as well as process noise, and thus increases the probability of finding a Trojan within a circuit. We also compare the method described in \cite{1} with the single parameter methods listed in papers \cite{2}, \cite{3},  \cite{4},  \cite{5}  and  \cite{6}.  

%\begin{flushleft}
\textbf{5.1 Methodology}
%\end{flushleft}

Every large integrated circuit has millions of transistors, and every transistor consumes some amount of leakage power when it is off, and also it contributes to the dynamic power consumption when it is on. This means that even a Trojan circuit when inserted in large IC will contribute to the overall static and dynamic power consumption. There are advantages of using Power analysis as a means to identify Trojans 1) It does not require the Trojan to be fully on to detect it,  and 2) It is non-invasive, meaning no extra circuitry is required to detect the Trojan.

The technique in this paper uses a multiple parameter power analysis method to detect Trojans. The technique in \cite{1} is compared with a transient power technique in \cite{4}, timing related techniques in \cite{3},\cite {5} and \cite {6} and transient current method given in \cite{2}.

However such non-invasive power techniques have a few challenges. 1) If the Trojan circuit is less than 0.1\% of the total circuit size it will cause a negligible change in the supply current and power consumption. 2) The impact of process variation on leakage current at lower design node can cause such a large variation that even a Trojan IC could emit a signature of a non-Trojan IC.

%\begin{flushleft}
\textbf{5.2 Multiple Parameter Trojan Detection}
%\end{flushleft}

In relation to \cite{1} the main issue with analyzing single parameters like I$_{DDT}$ and F$_{MAX}$ is process noise and process variation. Process variation can mask the effect of a Trojan and make it look like a genuine part of Circuit. If we plot the spread of I$_{DDT}$ with taking process variation into consideration, (Figure 1) we see that the Trojan is easily masked making it difficult to decipher. The Trojan in this example is an 8 bit counter and it has a very pronounced effect on I$_{DDT}$. This is a very serious issue because if a Trojan of a much smaller size is inserted into the circuit process variation will make it even more difficult to find the inclusion of a Trojan. If we take process noise into account the issue is even more serious, because process noise and process variation both help in the masking of the Trojan.

\includegraphics[width=3 in]{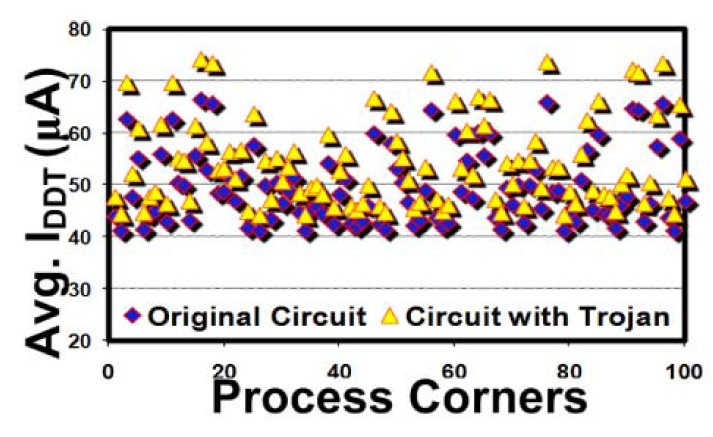}
\begin{center}
\textbf{Figure 1: Average I$_{DDT}$ versus Process Corners}
\end{center}

For this reason the inherent relationship between I$_{DDT}$ and F$_{MAX}$ is considered to form a multi parameter model to find the inclusion of Trojans. Consider Figure 2. In this figure Chip i and Chip j both have the same I$_{DDT}$ value but both chips are different because of the inclusion of the Trojan. But if you even more closely, both chips have a different F$_{MAX}$ value. This means both parameters contribute in finding a Trojan. Consider Chip i and Chip k, you can easily see that they have the same value of I$_{DDT}$. Under single parameter analysis both these chips would be considered one and the same. But with considering F$_{MAX}$ as well we can the difference between a genuine and a Trojan chip.

\includegraphics[width=3 in]{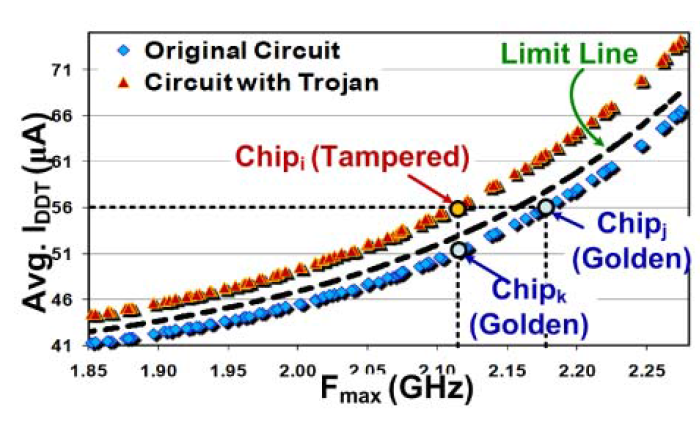}
\begin{center}
\textbf{Figure 2: Average I$_{DDT}$ versus F$_{MAX}$}
\end{center}

This was not possible with only one parameter analysis. Process variation and process noise can easily cause a lot of Chips which contain Trojans seem genuine(false positive) and a lot a genuine chips seem like infected chips(false negative). The intrinsic relation between I$_{DDT}$ and F$_{MAX}$ can be used to distinguish between chips under process noise. Now in the proposed approach F$_{MAX}$ is used to calibrating the test corners of the chip. It is calculated by taking the path of any one path within the chip. This makes it even more difficult for the adversary to insert a Trojan. A standard IC has more than 1,00,000 paths. It is very difficult for the adversary to guess exactly which path was used in calculating F$_{MAX}$. If the attacker even is able to guess which path was used to calibrate F$_{MAX}$ and introduces a Trojan in that path, it will cause a shift in the I$_{DDT}$ vs F$_{MAX}$  curve and the Trojan will still be detected. The only was an attacker can bypass the multi-parameter approach and alter F$_{MAX}$ is he/she can estimate the exact process variation, which is next to impossible to calibrate before the fabrication of the chip.

Another feature of the method in \cite{1} is to increase the sensitivity of finding a Trojan with taking process variation into account. The expression for sensitivity is given by 

%\begin{center} 
Sensitivity  = (I$_{tampered}$-I$_{original}$)/I$_{original}$ * 100
%\end{center}

From the expression we can see that, to increase sensitivity we have reduce I$_{original}$. 

\begin{flushleft}
\textbf{5.3 Test Vector Generation and Power Gating}
\end{flushleft}

In this approach we use Graph partition technique to divide the IC into different blocks and apply test vectors to each of these blocks independently. The blocks should be large enough to weed out process noise and should be small enough to eliminate background noise. Each partition should be independent because the test vectors are applied to each block individually. The MERO statistical approach is used in applying the test vectors. The test vectors are selected to increase the switching of the Trojan and hence increase I$_{tampered}$.

To improve I$_{original}$, a power gating mechanism is used. The power gating mechanism is used during test time to decrease the current from any other block. It is supplementary to the test vectors which are used in increasing the switching. We should remember that turning off all the remaining modules is not possible because some modules maybe interdependent. This would marginally increase I$_{original}$ but it still gives us much better sensitivity than without doing power gating. Hence using this technique we have a higher chance of finding a Trojan of a very small feature size.

The entire procedure can be summarized in Figure 3

\includegraphics[width=3 in]{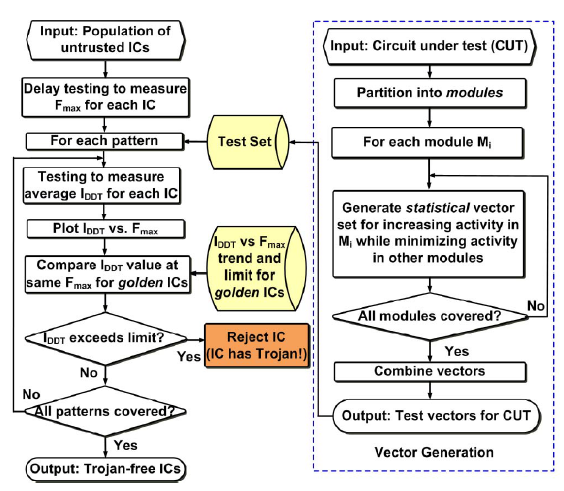}
\begin{center}
\textbf{Figure 3: Major Steps in Trojan Detection using Multiple Parameters}
\end{center}

\begin{flushleft}
\textbf{5.4 Other Side Parameters}
\end{flushleft}
We should note that other than I$_{DDT}$ and F$_{MAX}$ other parameters can also be used to characterize Trojans. Generally one parameter is affected by the Trojan in this case I$_{DDT}$ and the other parameter is used to characterize process noise(F$_{MAX}$). I$_{DDT}$ is the transient current of the circuit. Another component of the current is leakage current I$_{DDQ}$.Every transistor contributes to leakage. Hence even the Trojans are off they can contribute to leakage. Hence we have a relationship even with I$_{DDQ}$ and F$_{MAX}$. I$_{DDQ}$ monotonically increases with F$_{MAX}$.It shows the same behavior as I$_{DDT}$ vs F$_{MAX}$ .Hence any relation of I$_{DDT}$ with F$_{MAX}$ can be applied to I$_{DDQ}$ with F$_{MAX}$. Also I$_{DDT}$ and I$_{DDQ}$ are both input dependent. See Figure 4.

\includegraphics[width=3 in]{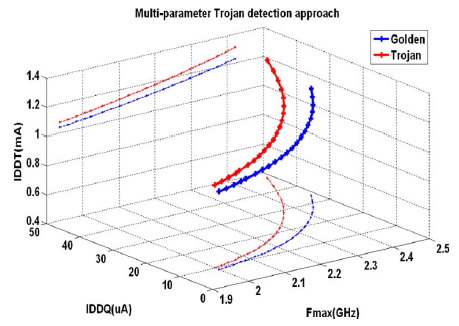}
\begin{center}
\textbf{Figure 4: The correlation among I$_{DDT}$ , I$_{DDQ}$, and F$_{MAX}$}
\end{center}

\begin{flushleft}
\textbf{5.5 Combination With Logic Testing}
\end{flushleft}

Side Channel Parameter testing may not be able to detect Ultra small Trojans due to various process variations and process noise, but are non-invasive. On the other hand Logic Testing techniques can detect Trojans with utmost confidence even with worst case process variation taken into account. But Logic Testing has a bad test coverage because most Trojans may have a certain input vector to trigger it. Hence it is inefficient as well. The authors in this paper \cite{1} suggest a method in which we can club the Multiple parameter analysis with Logic Testing and hence merging the best of both methods. Since Side Channel Analysis doesn’t require the Trojan to be fully on and Logic Testing being immune to process variation we  have a better chance in detecting a Trojan of lesser than 0.1\% size of the circuit. 

From the above lines we can draw a good conclusion to the method given in this paper. It gives us a brief relation between side channel parameters which can be used to find Trojans in a circuit. Many parameters other than the ones in this paper can be used. Moreover the Power gating and Test Vector generation methods can easily increase the sensitivity of the Trojan current. The consideration of process variation and noise is a good example of how scaling can help Trojans hide and multiple parameter analysis can find Trojans even with these variations taken into account.

Let us now compare the Multiple parameter analysis with a few single parameter analysis given in papers \cite{2} and \cite{4} and timing parameters in \cite{3},\cite{5} and \cite{6} see how it fares with these papers.

\begin{flushleft}
\textbf{5.6 Malicious Circuitry Detection using Transient Power Analysis for IC Security}
\end{flushleft}

The paper in \cite{4} deals with a Single Side Channel Parameter method for Detecting a Hardware Trojan. The authors have proposed a non-invasive Transient Power Scheme to detect the presence of a Trojan. They have also considered the effect of process noise and variations. The paper takes a statistical approach to find the presence of a Trojan.

All integrated circuits consume power. Power has to components Static and Dynamic power. When Switching occurs a transistor consumes dynamic power and in steady state operation it consumes static power. The same way, if a Trojan circuit is inserted , when it is a non-activated state it contributes to leakage(static) energy  and when an input vector activates it consumes dynamic power.
\begin{flushleft}
P = P$_{dynamic}$ + P$_{static}$\\
\end{flushleft}
  = (0.5*C*V$_{dd}$ $^{2}$  + Q$_{se}$*V$_{dd}$)*f*N + V$_{dd}$*I$_{leak}$
  \\
  \\
Taking all external factors into account we can summarize that power consumed by a chip is given by
Power consumed = Dynamic Power + Leakage Power + Measurement noise + Trojan Power

Trojan Power is only added to the power signature in presence of a Trojan. This is the way the power signature of an IC is calculated. 

The paper use the PCA (Principal Component Analysis) method to help in the Trojan Detection Scheme. The steps of the detection can be summarized as follows:\\
1) From a set of IC’s, select a few and Run Logical Tests on them to find their Power Signature.\\
2) Calculate the mean power of the sample set and subtract it from all the power traces to give the sample set a zero mean.\\ 
3) Find the Covariance matrix S of the population and perform Eigen vector decomposition on it and find its Eigen values and Eigen vectors. \\
4) Chose the top m Eigen vectors to form the transformation matrix A.\\
5) Use PCA to project the power traces of the genuine IC’s on the subspace of the Eigen Vectors. \\

Now for every untrusted IC we use PCA to generate its projections and project it on the Eigen vector subspace. Comparing the 2 spreads we can figure out if the IC is genuine or infected.

The paper is a very good example of involving statistical methods in the detection procedure. It involves a fast method of Trojan Detection for a big IC population. The Eigen vector projections are immune to process noise. It would work for a small process variation. The issue with this method is that since it’s a single parameter method it very immune to large process variations. The method could cause a lot of false positives or false negatives. Also there is a possibility that the Trojan won’t be activated at all. Hence the Trojan current would be much lesser than the IC circuit current. Hence the sensitivity in finding a Trojan is reduced. This would impact the Trojan detection procedure if the Trojan size is around 0.5 \% of the circuit size or lesser. The Multi-Parameter method in \cite{1} 

Addresses these issues with 2 parameters in which 1 is affected by process noise and the other parameter is affected by process variation.

\begin{flushleft}
\textbf{5.7 Hardware Trojan Detection Using Path Delay Fingerprint}
\end{flushleft}

So far we have seen papers which deal with Power Signatures to detect Trojans. The paper in \cite{3} discusses a method to use a Path Delay to detect a Trojan in a circuit. The paper talks about implicit and explicit payload Trojans. A Trojan when activated in a circuit can change the delay timing of the paths within the circuit. This delay can be affected by process variation, but with proper filtering can be used as method of Trojan detection.

The method discussed in this paper can effectively detect Explicit Payload Trojans. Explicit payload Trojans can alter internal signals of an IC. Hence they add to the delay of the signals. Implicit Payload Trojans do not affect any signal in the circuit. They use the existing signals to generate their own output signals. Hence they do not add to the delay.
 
The basic procedure can be described in 3 steps, which are as follows:\\
\\
Path Delay Gathering from a nominal chip:\\
\\
For this step, High coverage test patterns are given at the input and the path delay information is extracted. Process variation can play a big part in deep-submicron technology and can vary as much as 5\%. In this paper the authors vary the delay parameters of the SMIC technology library by 7.5\% in both directions. A Logic Synthesis software is then used to compile the test circuit under consideration to give its gate level netlist without a Trojan.  Trojans are then inserted into the netlist. The modified netlist is then passed through a Static Timing Analysis tool to generated the SDF(Standard Delay Formats)  Files for every circuit under consideration. The SDF contains the path delay for all paths in the Test Circuit.\\
\\
Delay Testing:\\
\\
Test Patterns should be conclusive to find out all the characteristics of a Circuit under test. Hence a lot of Test Patterns should be generated for this case. The paper mentions using a ATPG tool which analyses any input netlist and generates the set of input patterns. The input patterns are then applied to the chip to get its corresponding output vectors. The numbers of path delays for any chip under consideration can be roughly calculated as the number of inputs*number of outputs, which is a very big test vector.\\
\\
Sample Analysis and Trojan Detection:\\
\\
For detecting a Trojan, analysis is done of the entire data set. Since the data set has a lot of dimension it has to be reduced before we can actually deal with the data. A PCA method can be used to reduce the number of dimensions. However a PCA analysis can cause a loss of important data in the dataset. Hence the data is first divided into smaller sets and then a PCA is done to reduce the dimensionality.

For detecting the Trojans a Quickhull algorithm is used to construct a convex hull of a set of points. A convex hull is the smallest convex set which contains all the points. This can be used in Trojan Detection. The PCA causes a reduction in dimension and the by using diving the test vector into smaller units we can use these sub units to construct their corresponding convex hulls (Figure 5).

\includegraphics[width=3 in]{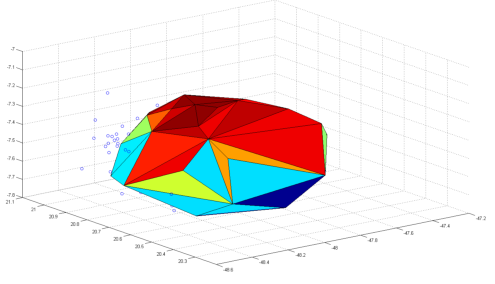}
\begin{center}
\textbf{Figure 5: Convex Hull of Test Vector Space}
\end{center}

The number of dimensions of the convex hull is reduced to 3 to give a uniform three-dimensional space to plot the points. From Convex Hull theory we can come to the conclusion that the points which nearer to the surface of the Convex Hull are genuine and the ones which are further from the convex hull are termed as Trojans.

The authors of this paper have tested this method with various Trojans including both Explicit and Implicit Payload Trojans and moderate to worst process variation corners. The Trojans implemented by them range from 2\% to 0.76\% of the total chip area. They have reported a 100\% hit rate for Explicit Trojans and a 36\% hit rate for the Implicit variety.

The paper has made a good attempt to include worst case process variation into account while doing analysis. The authors have also conducted tests in which they have introduced the Trojans in various parts of the circuit and have achieved good results. The only issue with this method is the ineffectiveness to find Implicit Payload Trojans which could be used to Transmit Information but do not affect timing statistics of the chip. The Multi –Parameter method in \cite{1} should be able to find both Implicit and Explicit Payload Trojans since the type of Trojan does not matter in its method of detection.

\begin{flushleft}
\textbf{5.7 At-Speed Delay Characterization for IC Authentication and Trojan Horse Detection}
\end{flushleft}

The paper in\cite{5} deals with a Run-Time method to predict the presence of a Hardware Trojan. It uses timing information of the combinational paths to find out if a Trojan is present or not.  Currently many existing techniques like Physically Unclonable Functions (PUF) circuits are used to authenticate IC’s. PUF’s have challenge response architecture to Authenticate IC’s. This normally consists of Interconnected Flip-Flops which accept an n-bit challenge vector. 

The number of Flips-Flops are synonymous to the number of test Inputs. For each n-bit input there is 1 bit output which is very inefficient and can be re-engineered by the attacker with some effort. There is also an Area overhead and Speed issue with the number of test patterns increasing.

The paper introduces another run-time method to predict the presence of a Trojan using a Timing Signature. The method uses a Delay Measurement method At-speed which does not affect the circuit functionality. The basic motivation for this paper is that, when a Trojan is activated the Delay Characteristics of circuit change. While PUF’s deal with non-functional parts of the circuit, the technique in this paper deals with functional parts of the circuit as well as the non-functional parts.

The Path Delay Characterization method uses negatively skewed registers which are placed in the combinational path within the circuit. The architecture consists of the Main Circuit – one register to register combinational path delay, clk1 – which is the main clock in the system, Shadow Register – a negatively skewed register and the Destination Register – The final register in the combinational path.

The Shadow register takes the same input as the Destination register. The Shadow register is clocked by clk2 which is negatively skewed with respected to clk1.This means the Shadow Register receives the input before the Destination Register. The contents of both the registers are compared after every clock cycle. If there any discrepancy, there is a 1 bit Result bit which is set to 1 when there is a difference. The clk1 remains constant but the skew in clk2 is changed in steps and the results in both the Registers are compared for various values of skew. If the results remain consistent the Result bit remains 1. Now no other Register input can change the value of the Result bit after it is set to 1.The results of the various registers are then read through a scan chain. The presence of a 1 in the chain confirms the presence of the Trojan. If we back annotate to the Destination register which was set to 1 we can predict the possible location of the Trojan.

The authors implemented this scheme of Virtex-II FPGA. The FPGA’s DCM (Digital Clock Manager) was used to change the skew of the clock by providing a phase shift in the clock. The total time for one delay measurement process is given by s*(c+p) , where, s – number of skew steps, c – number of clock cycles in each skew steps and p – number of paths the delay is measured. This means the delay measurements will be performed every s*(c+p) cycles. The method also incorporates an on-chip Ring Oscillator which is used a Temperature Compensation Network. The Delay of the registers can change due to temperature effects. This mechanism makes sure that temperature irregularities can be compensated.

The verification mechanism consists of Authentication Control which generates the test inputs. A signature is also present on-chip which gives the desired response to the given test vector. The response is calibrated by calculating the path delay of each of the individual paths of the circuit. Since this an inherent circuit feature it can be regarded as the timing signature of the circuit under test. The delay is calculated in the following manner- The skew of clk2 is increased in a number of steps till a 1 is generated. The skew which generates the 1 is then subtracted from clk1 to give the delay of the circuit.

The paper gives a good account of how Trojan Detection can be done at runtime. It only includes minimum hardware of 1 register per path which is not a very big area overhead. It can help the IC authenticator to detect the presence of a Trojan after the chip is received from the foundry. The technique is non-invasive and the only changes done to the Design are done before it is sent to the foundry.

The paper has a few cons primarily delaying with skew and process variation. It is very difficult to have control of on chip using a clock source without taking into account power fluctuations and other secondary effects. Also having a dual clock in a circuit is unfeasible because of various clocking issues. The main worry is if the adversary understands the design in the foundry and selectively disables the Shadow Register mechanism. They can even make changes so that an error is never reported. Our primary paper in \cite{1} takes process variation into account and also deals with a single clock. It is also immune to jitter and other factors and also doesn’t add any design overhead.

\begin{flushleft}
\textbf{5.7 RON - An On Chip Oscillator Network for Hardware Trojan Detection}
\end{flushleft}

A Ring Oscillator is a circuit which has an odd number of gates and is used to set the frequency of a circuit. It can be a set of inverters or a set of any combinational gates. The authors of this paper \cite{6} suggest the usage of such a structure in helping find a Trojan. The paper mentions that this structure effectively eliminates the issue of measurement noise, helps measuring dynamic power, and additionally compensates for the impact of process variations. Using statistical methods the detection of Trojans is possible.

The principle of this technique is that power signature of an Infected IC will be different from a Genuine one. A Ring Oscillator Network has the ability to detect power fluctuations. A number of RON can be placed around the circuit which can be used as power monitors. The frequency of this oscillator is given by the delay of the all the gates (inverters) in the chain. For a n stage ring oscillator the delay is 2*n*td and the frequency is given as 1/(2*n*td). The delays of the inverters are impacted by various process variations as well as Power Supply Variation. When the voltage across the chip drops the delay of the inverter increases.

If a Trojan is inserted in an IC, the switching gates in the Trojan would cause a small voltage drop on the power lines. The power supply noise for Trojan free and Trojan infected IC’s would be different. Using this fact in mind a Trojan IC can be detected. Since one RON will be insufficient for a chip, multiple Ring Oscillator Networks are inserted all around the chip to capture the power supply and noise effects. The outputs of the Ring Oscillator Network can be used as a power signature to distinguish between free and infected IC’s. The Oscillation count from RON is used in generating the power signature. The formula for Oscillation count is given as follows:

\[C_i= \int_0^T \frac{1}{2*n*t_{di}(t)}\,dt.\]

Using the difference of cycle count between the genuine IC and Trojan IC the distance between and genuine and infected IC is generated. A circuit with multiple RON’s uses a multiplexer to select the RON to do authentication and another multiplexer selects another RON to do the recording.

To deal with process and other variations statistical methods such as Single Outlier analysis and Principal component Analysis are used. If Oscillation Cycles of a RON lie within a certain threshold the IC is determined as genuine. Since the number of Ring Oscillators  used within the IC are large in number, with many Oscillation counts, the PCA method is used to reduce the number of dimensions to deal with.

A Convex Hull is then constructed with the top 3 components. If output of the RON lies beyond the Hull, the IC
has failed authentication. IC’s which have passed authentication are further analyzed using Advanced Outlier Analysis. For experimental analysis the authors tested 10 different Trojans whose size ranges from 0.36-0.9\% of the circuit size. The average Trojan detection rate was 80-90\%.

From our analysis of the paper we can say that the paper describes a very good way inclusive of process variation to detect Trojans. Also in layout the Trojan has a distributed nature which makes it difficult to disable. Another fact is that multiple RON networks give a good coverage of the IC layout. The only Cons that we could think of the effect of jitter because not all the RON’s would suffer from the same amount of jitter due to their distributed placement. Also multiple RON’s can have unnecessary area overhead if IC’s are of medium size. The multi-parameter technique in \cite{1} does not add any area overhead and also take process variation into account. Even the effect of skew and jitter are effectively modeled in this procedure.

\begin{flushleft}
\textbf{5.8 Sensitivity Analysis to Hardware Trojans Using Power Supply Transient Signals}
\end{flushleft}

The paper in \cite{2} uses Transient Power as an Electrical Signature to detect Trojans. Transient Power is generated by Transient Current which is caused by the switching of gates. The analysis is done at multiple port to improve the resolution of Trojan Detection. There could be an attacker who could design a Trojan in such a manner that it has minimal impact on I$_{DDT}$ and  I$_{DDQ}$. Also background noise can help in masking the whereabouts of the Trojan.

This method is a statistical method and needs a simulation model. It analyses an IC’s supply current measured from multiple supply ports to deal with small Trojan-signal-to-background-current ratios. A simple calibration procedure is used to reduce the impact of process variations and background noise. Simulation models are extracted from a layout under test and the data is analyzed to detect the possibility of a Trojan.

For experimental analysis, Trojans were manually inserted in the interstitial places in the layout. The Trojan is a special set of gates which has a constant value on one of its inputs and the second input is derived from the node in the layout. The I$_{DDT}$ signals are measured from various ports on the chip. A test sequence is applied to cause switching in the gates.

The Calibration procedure deals with mitigating the effect of process variation which occurs on the chip core, power grid and off chip connections. A p channel transistor with its gate connected to scan FF and its source connected to power port of the chip is used in the calibration. The scan FF generates a step input which triggers the p transistor and causes a short between V$_{DD}$ and Gnd. The impulse response are generated from the step matrices and inserted in a calibration matrix. Using the calibration matrices from the various ports a Transformation matrix is generated.\\
\\
Trojan Detection Analysis:\\
\\
A scatter plot is used for statistical analysis. Two separate test sequences are applied and I$_{DDT}$ is measured from the various ports. Calibration is conducted to reduce process variation. The calibrated results are then plotted on scatter plots for analysis. The mean and variance of the scatterplot is used to determine the statistical limits of the enclosing ellipse. The enclosing ellipse defines a set of points which a Trojan Free to reside within the ellipse and which are infected to reside outside it. If any point lies outside an Outlier Analysis is performed to ascertain if a Trojan is present or not.

The paper uses a calibration method to remove the effect of process variation which is very important with increased scaling. As the scatter plot analysis method help in fast yet accurate method in analyzing current and power traces. The technique could be extended to other side channel parameters like timing or leakage current.
The only issue is that the equipment used in the calibration procedure itself could be immune to process variation which could lead the mitigation process having some residues. The multi-channel parameter technique in \cite{1} mitigates this calibration issue by having 2 parameters, in  which one is affected by process variation only and the other is affected by process noise only, which has a better tendency to reduce process variation.
\section{Results}
\parskip 0.1in
Table 1 shows the Trojan Coverage using only MERO and using both a combination of MERO and the conventional multi side channel method. We see that for the given benchmarks an average of 99.98\% of the Trojans were detected. The size of the Trojans varied from 2 to 16 inputs. Hence an ultra-small to small Trojans have a good detection accuracy using the combination of MERO and the conventional multiple parameter side channel
method for Trojan insertion.

\includegraphics[width=3 in]{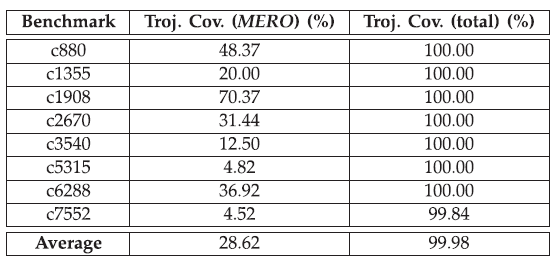}
\begin{center}
\textbf{Table 1: Trojan Coverage for ISCAS-85 Benchmark Circuits}
\end{center}
\section{Conclusion}
\parskip 0.1in
For the purpose of this survey we have reviewed six papers related to Hardware Trojans. The basic issue which was understood during the analysis was that process variation and process noise play an important role in hindering the search of Trojan in an IC. Most Single Parameter methods can only take care of either noise or variation, but not both. Hence we see that they suffer from a limited accuracy for ultra-small Trojans which are about 0.02\% of the entire circuit size. The main paper \cite{1} which we reviewed had a very novel approach which two side channel parameters into account for the Trojan Detection. We see not only does this increase the detection resolution but also reduces the complexity of the test vector space. With the combination of Logic Testing and Multi0Side Channel analysis the detection rate can be increased further.

\section{Critique}
\parskip 0.1in
\textbf{Pros} 
\begin{itemize}
 \item The paper revealed a very novel idea using the combination of 2 interdependent parameters.
  \item The integration with Logic Testing is good way to club 2 different Trojan detection methods.
  \item Selective Test Vector generation and Power Gating not only increases detection sensitivity but also reduces the complexity of test vectors to activate the Trojans.
  \item Overall the paper was a very good read and used a simplified language and always kept the reader engaged.
  \item Process Variation was covered in great detail in this paper.
\end{itemize}
 
\textbf{Cons} 
\begin{itemize}
  \item The power gating mechanism actually adds a big area overhead and also causes a modification in the original circuit which requires specialized tools for the implementation.
  \item The integration with Logic Testing will increase the testing time by 2-3 times which should have been mentioned in the paper.
\end{itemize}

%%%%%%%%%%%%%%%%%%%%%%%%%%%
% References %
%%%%%%%%%%%%%%%%%%%%%%%%%%%
\small
\balance
\bibliographystyle{abbrv}
%\bibliographystyle{unsrt} %Entries are not ordered alphabetically, but in the order they are first referenced.
%\bibliography{reference-list}
%\bibliography{../bib/diversity,../bib/energy-all, ../bib/diversity }

\begin{thebibliography}{}

\end{thebibliography}


\begin{thebibliography}{100}  % 100 is a random guess of the total number of 
%references
\parskip 0.1in

\bibitem {1} S. Narasimhan, D. Du, R. S. Chakraborty, S. Paul, F. G. Wolff, C. A. Papachristou, K. Roy and S. Bhunia. Hardware Trojan Detection by Multiple-Parameter Side-Channel Analysis. In \emph{IEEE Transactions on Computers Vol.62 No.11 November 2013}.

\bibitem {2} R. Rad, J. Plusquellic and M. Tehranipoor, “Sensitivity Analysis to Hardware Trojans using Power Supply Transient Signals”.

\bibitem {3} Y. Jin and Y. Makris, “Hardware Trojan Detection Using Path
Delay Fingerprint,” Proc. IEEE Int’l Workshop Hardware-Oriented Security and Trust, pp. 51-57, 2008.

\bibitem {4} L. Wang, H. Xie and H. Luo, "Malicious Circuitry Detection Using Transient Power Analysis for IC Security", 
QR2MSE 2013.

\bibitem {5} J. Li and J. Lach, "At-Speed Delay Characterization for IC Authentication and Trojan Horse Detection"

\bibitem {6} X. Zhang and M. Tehranipoor, "RON: An On-Chip Ring Oscillator Network for Hardware Trojan Detection", EDAA 2011

\end{thebibliography}

%\input{appendix.tex}

\end{sloppypar}
\end{document}